\documentstyle[12pt]{article} 
\newcommand{\be}{\begin{equation}}
\newcommand{\ee}{\end{equation}}
\newcommand{\ul}{\underline}
\newcommand{\ba}{\begin{eqnarray}}
\newcommand{\ea}{\end{eqnarray}}
\begin{document}

\begin{center}
{\bf Trilinear generally covariant equations of {\rm AP}}
\end{center}
\centerline{ I.\ L.\ Zhogin\footnote{%
e-mail: zhogin@inp.nsk.su}
}
\hspace{2cm}
\\
\\

Field equations for $n$-frames $h^a{}_\mu$ that are possible in the 
theory of absolute parallelism (AP) are considered. 
The methods of compatibility (or formal integrability) theory
enable us to find the non-lagrangian equation having unusual kind
of compatibility conditions, guaranteed by two (not one) identities. 
This 'unique equation' was not noted explicitly in the classification by 
Einstein and Mayer of compatible second order AP equations.

In this work it is shown that some equations of AP (including
'unique equation') can, after the substitution 
$h^a{}_\mu= H^p H^a{}_\mu$ ($H= \det H^a{}_\mu$,
 $p$ is an $n$-dependent constant), be written in a trilinear form that
 contains only the matrix $H_a{}^\mu $ and its derivatives and not 
inverse  matrix $H^a{}_\mu$. 
 The equations are still regular 
(and involutive) for degenerate but finite matrices $H_a{}^\mu$ 
if rank$\, H_a{}^\mu \geq 2$.

\hspace{2cm}
\\
\\

    1.  In this paper, we consider the theory of a Riemannian space with
 absolute parallelism proposed in [1-3]. 
A number of modern works (see  [4,5] and references therein)
 deal with lagrangian equations
 (this class of equations is only a small part of
 Einstein--Mayer (EM) classification
of compatible second order equations of AP [1]),
but we want to show that it is not the most interesting of AP.

The geometry of AP is specified by a field of
 coframes
\[
h^a{}_\mu \in GL(n), \]
which may be exposed (is determined up) to a global rotation:
\be \label{lor}
h^{*a}{}_\mu(x)= s^a{}_b\, h^b{}_\mu(x),
\mbox{ where } s^a{}_b =\mbox{const;}  
\ s^a{}_b\in O(1,n-1)\ ;
\ee
\[ \mbox{i.e.,\ \ }
  \eta_{ab}s^a{}_c\, s^b{}_d =\eta_{cd}\, ,  \mbox{ where \  }
 \eta_{ab}=\eta^{ab}=\mbox{diag}(-1,1,\ldots,1) \,  
\mbox{ (Minkowski metric)}. \]

 With respect to coordinate transformations, the Latin and Greek indices
 have scalar and vector natures, respectively
 (for Lorenz transformations (\ref{lor}), all is quite the contrary). 
We introduce the metric
\be
g_{\mu \nu }= \eta_{ab}h^a{}_\mu h^b{}_\nu
=h^a{}_\mu h_{a\nu} \ee
and the usual covariant differentiation with the symmetric connection. 
We also define the tensors (differential covariants)
\be
\label{gal}
\gamma_{a\mu\nu} = h_{a\mu;\nu}\ , \ \ \
\Lambda_{a\mu\nu} = 2\gamma_{a[\mu\nu]}=h_{a\mu,\nu}- h_{a\nu,\mu}\ .
\ee

In covariant expressions, we shall omit in contractions the matrices
\[
g^{\mu\nu} \mbox{ and } 
\eta^{ab}  \mbox{ (since }
g^{\mu\nu}{}_{;\lambda}=0, \ \ \eta^{ab}{}_{;\lambda}=0),  
\]
understanding
\be
\Psi_{\cdots a \cdots  a \cdots } =
\Psi_{\cdots  a \cdots  b \cdots } \eta^{ab}\ ,
\ \ \ \ \Psi_{\cdots \mu\cdots \mu\cdots } =
\Psi_{\cdots   \mu \cdots  \nu \cdots }  g^{\mu\nu}\ . \ee
The type of an index is changed by means of $h^a{}_{\mu}$;
 the notation (tensor identification) is unchanged (except for  
$h_{a\mu},\ g_{\mu\nu},\ \eta_{ab}$), 
for example:  
\[
\Lambda_{\mu\nu\lambda}= h_{a\mu} \Lambda_{a\nu\lambda}.  
\]

We introduce also the notation ('scalar differentiation'
 and irreducible parts of $\Lambda$):
\be  \Psi_{a\cdots \mu,b} = \Psi_{a\cdots \mu ;\nu}  h_{b\nu}\,; \ee
\begin{equation}
\label{sde}
\Phi _a =\Lambda_{bba};\ \ 
S_{abc}=3\Lambda_{[abc]}=\Lambda_{abc}+ (abc)
=6\gamma_{[abc]}~;
\end{equation}
\be \label{fde}
f_{ab}=\Phi_{a,b}-\Phi_{b,a}+\Phi_{c}\Lambda_{cab} \
\mbox{ or \ } 
f_{\mu\nu}=2\Phi_{[\mu;\nu]}=\Phi_{\mu,\nu}-\Phi_{\nu,\mu} \ . \ee
Note that
\be \Lambda_{abc} =-\Lambda_{acb}; \ \ 
\gamma_{abc}=-\gamma_{bac}=1/2\,S_{abc}-\Lambda_{cab}.
\ee
For the transposition of scalar indices of differentiation
 there is a simple rule (here, $\Psi$ is a scalar, 
i.e., has only Latin indices):
\be 
\Psi_{a\cdots b,c,d} - \Psi_{a\cdots b,d,c} 
= - \Psi_{a\cdots b,e} \Lambda_{ecd} \, .
\ee

Sometimes, transition to Greek or mixed indices makes an expression
 more evident and plain; for example,
 from Eq.\ (\ref{gal}) there follows the well-known identity [1]:
\be
\Lambda_{a[\mu\nu;\lambda]} \equiv 0~ 
\mbox{ \ \ or \ }
\Lambda_{abc,d}+\Lambda_{ade}\Lambda_{ebc}
+ (bcd) \equiv 0~.
\ee
Contracting two indices, we obtain [see (\ref{sde}), (\ref{fde})]:
\begin{equation}
\label{ide}
\Lambda_{abc,a}+f_{bc}\equiv 0 \ .
\end{equation}

2. The equations of the frame field can be expressed in the form 
(separately symmetric and antisymmetric parts; 
see  (\ref{gal})--(\ref{fde}), [6]):
\ba
\nonumber
     {\bf G}_{\mu\nu} & = & 
2\Lambda_{(\mu\nu)\lambda;\lambda}+\sigma
(\Phi_{\mu;\nu}+\Phi_{\nu;\mu}-2 g_{\mu\nu}
\Phi_{\lambda ;\lambda })+ (\Lambda^2) \\
\label{sya}
 &  = & -2G_{\mu\nu}+(2\sigma-2)(\Phi_{(\mu;\nu)}-g_{\mu\nu}
\Phi_{\lambda;\lambda})+U_{\mu\nu}(\Lambda^{2})=0, \\
\label{syb}
{\bf H}_{\mu\nu} & = & S_{\mu\nu\lambda;\lambda}+\tau f_{\mu\nu}
+V_{\mu\nu}(\Lambda^2)=0~.
\ea
Here $\sigma$ and $\tau$ are certain constants; 
$G_{\mu\nu}$ is Einstein tensor ($G_{\mu\nu;\nu}\equiv 0$); 
$U_{\mu\nu}=U_{\nu\mu},\ \ V_{\mu\nu}=-V_{\nu\mu}. $.

With regard to questions of compatibility (or formal integrability)  
of non-linear systems of partial differential equations, 
the reader is referred to [7]:  \\
{\em Corollary 4.11.} \
 The system ${\cal R}_{\rm q}$ is involutive (and compatible) 
if its symbol
$G_{\rm q}$ is involutive and the mapping 
$\pi^{\rm q+1}_{\rm q}:\  {\cal R}_{\rm q+1} \to 
{\cal R}_{\rm q}$ 
is surjective.

 If we differentiate the second-order system ${\cal R}_{2}$ 
[see Eqs.\ (\ref{sya}), (\ref{syb})] 
and see that in a certain combination of equations 
(from system ${\cal R}_{3}$) the higher derivatives ($h'''$)
 cancel, then if
 $\pi^{3}_{2}:  {\cal R}_{3} \to {\cal R}_{2}$ 
is to be surjective the remaining terms must also cancel by means of
${\cal R}_{2}$
(otherwise a new second-order equation appears and it is, moreover, 
irregular in first jets), i.e., a corresponding identity must exist.

The symbol $G_2$ (vector space family over ${\cal R}_{2}$) 
of the system ${\cal R}_{2}$ can be determined by the linear system
\begin{equation}
\label{smb}
G_2:\ \  {\bf e}_{ab}=\frac{\partial ({\bf G}_{ab}+{\bf H}_{ab})}
{\partial h_{c\mu,\nu\lambda}}u_{c\mu,\nu\lambda }=0,\ \ 
(u_{c\mu,\nu\lambda }= u_{c\mu,\lambda \nu}).
\end{equation}
(In coordinates $u_{a\mu,\nu\cdots \tau}$ indices following 'comma' 
(here 'comma' does not mean coordinate derivative)
are symmetric with respect to transpositions: 
$u_{a\mu,\nu\cdots \tau}= u_{a\mu,(\nu\cdots \tau)}$).

The involutory property of the symbol $G_2$ must be verified 
over all points $(x, h, h', h'')$ in ${\cal R}_{2}$, 
but Eq.\ (\ref{smb}) 
contains only the matrix $h^a{}_\mu$ 
(and inverse matrix $h_a{}^\mu$), 
which can be made equal to the unit matrix 
by coordinate transformation 
provided $h^a{}_\mu$ is non-degenerate. 
If $h^a{}_\mu = \delta^a_\mu$, then Eq.\ (\ref{smb}) 
is closely similar to linearized equations.

We need also {\em Definition 2.16} of [7]:
the symbol $G_{\rm q}$ is involutive if
\begin{equation}
\label{inv}
\dim  G_{{\rm q+1}}= \dim G^0_{\rm q}
+\dim G^1_{\rm q} + \cdots
+\dim G^{{\rm i}}_{\rm q} + \cdots
+ \dim G^{n-1}_{\rm q}\ \ (\dim G^n_{\rm q}\equiv 0) .
\end{equation} 

For analyzing symbols, it is more convenient to use 
the Euclidean signature and indices $a,\mu =1, \ldots, n$ 
(while $0\leq{}$i${}\leq n$).
 
The subspaces  $G^{{\rm i}}_2$ are determined by addition 
to Eq.\ (\ref{smb}) of
 the following equations:
\be
\label{gi2}
u_{ab,cd} = 0,
\mbox{ \ if \ } c\leq  {}{\rm i}   \mbox{ \ or \ } d\le {\rm i}
  \ (G^{0}_{2}=G_{2}).
\ee
These equations are non-covariant (with respect to infinitesimal 
coordinate diffemorfisms and Lorenz transformations), 
but involutory property of symbol is covariant one [7], of course. 

We turn to Eqs.\ (\ref{sya}) and (\ref{syb}).  The equation  
${\bf G}_{\mu\nu;\nu} =0$ gives $f_{\mu\nu;\nu}=J^{(1)}_\mu$, or in
Latin indices (for the interest of the thing)
\be \label{j1}
{\bf G}_{ab,b} = (\sigma-1)[f_{ab,b}-\frac12\Lambda_{abc}f_{bc}
+f_{ab}\Phi_b - J^{(1)}_a(\Lambda'\Lambda, \Lambda^3)]=0.
\ee
From the equation ${\bf H}_{\mu\nu;\nu} =0$ there follows the analogous
(``Maxwell'') equation
\be \label{j2}
{\bf H}_{\mu\nu;\nu} =\tau(f_{\mu\nu;\nu} -J^{(2)}_\mu )=0.
\ee
It is clear that (if $ J^{(1)}_\mu = J^{(2)}_\mu =J_\mu$) the combination 
\[
\tau{\bf G}_{\mu\nu;\nu}+(1-\sigma){\bf H}_{\mu\nu;\nu}
\]
must become identity when Eqs.\ (\ref{sya}) and (\ref{syb})
are taken into account:
\begin{equation}
\label{i1}
\tau{\bf G}_{ab,b}+(1-\sigma){\bf H}_{ab,b}\equiv
A_{abc}(\Lambda) {\bf G}_{bc}+ B_{abc}(\Lambda) {\bf H}_{bc}.
\end{equation}
Here $A$ and $B$ are linear in $\Lambda$.

It can be shown that the symbol $G_2$ is involutive if 
$\tau\neq 0$ ($\forall \sigma$), and therefore the system 
(\ref{sya}), (\ref{syb}) will be 
compatible in the presence of identity (\ref{i1}).

For $\sigma=1$ (only the symmetric part occurs in identity) we can 
have the equation of vacuum general relativity (VGR):
\[
-\frac12 {\bf G}_{\mu\nu} = R_{\mu\nu} -\frac12 g_{\mu\nu} R =0,
\]
and in this case skew-symmetric part (\ref{syb}) can be arbitrary 
(if $\tau\neq0$).

We come now finally to the most interesting case (details in [6]):
\ $\tau=0,\ \sigma\neq 1$.
Identity ({\ref{i1}) contains only the antisymmtric part, and for it 
there is a unique possibility:
\be \label{sb}
{\bf H}_{\mu\nu} = S_{\mu\nu\lambda;\lambda} =0; \ \ 
{\bf H}_{\mu\nu;\nu} \equiv 0.
\ee
The symmetric part cannot be chosen arbitrary; for one must ensure 
that the equation
\be
J_{\mu;\mu}=0 \ (J_{\mu}= J^{(1)}_{\mu}, \mbox{\ see Eq.\ (\ref{j1})})
\ee
becomes an identity.

The symbol $G_2$ of the system (\ref{sya})--(\ref{syb}) 
for $\tau=0,\ \sigma\neq 1$ is not involutive, 
but its prolongation symbol $G_3$ is,
and for compatibility it is necessary (and sufficiently) 
to ensure also the second identity\footnote{It is very beautifully.}.

It can be shown [6] that there exists a unique 
compatible system with  $\tau=0$ (and $\sigma\neq 1$), 
for which $\sigma=1/3$ and
\be
\label{me}
f_{\mu\nu;\nu} = (S_{\mu\nu\lambda}\Phi_\lambda){}_{;\nu}\ .
\ee
It is convenient to express this unique equation of AP 
(in some sense, it is antipode of VGR in the set of AP  equations) 
in the form
\be
\label{s1}
\frac12({\bf G}_{a\mu}-{\bf H}_{a\mu})
=L_{a\mu\nu;\nu}- \frac13
(f_{a\mu }+L_{a\mu \nu }\Phi _{\nu })=0\ \ (\sigma=1/3,\ \tau=0),
\ee
where
\be \label{ld}
L_{a\mu \nu }=-L_{a\nu \mu}=\Lambda_{a\mu \nu }-S_{a\mu \nu }-
1/3( h_{a\mu }\Phi_{\nu } - h_{a\nu }\Phi_{\mu }) ~.
\ee
    
The trace equation ${\bf G}_{aa}=0$ [see (\ref{sya}), 
(\ref{s1}), (\ref{ld})] becomes irregular for $n=4$:
\be
{\bf G}_{aa}= \frac23 (4-n) \Phi_{aa} + Q(\Lambda^2)=0,
\ \ Q \not\equiv 0\, .
\ee
Therefore, additional spatial dimension(s) is needed.

    3.  We consider one further AP system 
(it looks simple in Latin indices)
\be \label{s2}
\frac12({\bf G}_{ab}+{\bf H}_{ab})=
\Lambda_{abc,c} + \Lambda_{acd}\Lambda_{cdb} =0
\ (\sigma=0,\ \tau=1),
\ee
that, like Eq.\ (\ref{s1}), admits solutions without 
"electromagnetic field", i.e., remains compatible 
on addition of the equation
\be \label{f0}
f_{\mu\nu} =0 \, .
\ee
System (\ref{s2}) was considered in [3], together with (\ref{f0}).

In general case, the tensor $V_{\mu\nu}$ in Eq.\ (\ref{syb}) 
can contain three terms:
\be \label{v3}
V_{\mu\nu}= a_1 S_{\mu\nu\lambda}\Phi_\lambda +
a_2 \Lambda_{\lambda \mu\nu }\Phi_\lambda +
a_3 ( \Lambda_{ \mu\varepsilon\tau }\Lambda_{ \varepsilon\tau \nu }
- \Lambda_{ \nu\varepsilon\tau }\Lambda_{ \varepsilon\tau \mu } ).
\ee
If $a_2=0$ and $a_3=0$, the system remains compatible 
on the addition of Eq.\ (\ref{f0}). 
Indeed, although the irregular equation
$J_\mu=0$ follows from Eqs.\ (\ref{j2}) and (\ref{f0}), 
it becomes an identity when (\ref{syb}) and (\ref{f0}) 
are taken into account:
\be
J_\mu =J_\mu^{(2)} \sim (S_{\mu\nu\lambda}\Phi_\lambda)_{;\nu}
= -\frac12 S_{\mu\nu\lambda} f _{\nu\lambda} + 
\tau  f_{\mu\lambda}\Phi_\lambda\, .
\ee

For system (\ref{s2}), the symbol $G_2$ is determined by the
equation (see (\ref{smb}); $h^a{}_\mu =\delta^a_\mu$)
\be \label{sm2}
{\bf e}_{ab}= u_{ab,cc} - u_{ac,cb} =0,
\ee
(it differs from the Maxwell equation 
$ A_{b,cc} - A_{c,cb} =0$ only in the "redundant" index $a$).

We give the values of 
$\dim G_3$ and $\dim G_2^{\rm i}$ for system (\ref{s2}):
\ba \label{g2}
\dim G_2^{\rm i} & = & n^2(n- {\rm i})(n- {\rm i}+1)/2 - n^2 + 
n^2 \delta ^{\rm i}_n
+ n \delta ^{\rm i}_{n-1}\, ;\\
\label{g3}
\dim G_3  & = & n^3(n+1)(n+2)/6 - n^3 + n \, .
\ea
In  (\ref{g2}), the first term expresses the number of coordinates
$u_{ab,cd}$
with allowance for Eq.\ (\ref{gi2});
the second, the number of equations (\ref{sm2});
 and the third, the fact  that for $G_2^{\rm n}$ 
all equations (\ref{sm2}) become an identity in view of (\ref{gi2}).
Finally, the last term $n \delta ^{\rm i}_{n-1}$  is added, 
since for  $G_2^{{\rm n} -1}$  $n$ equations in (\ref{sm2})  
become an identity:
\be
{\bf e}_{a{\rm n}}= u_{a{\rm n},{\rm nn}} - u_{a{\rm n},{\rm nn}} 
\equiv 0\, ;
\ee
here {\rm\large n} means 
the fixed value of the index; no summation over it.

Terms of Eq.\ (\ref{g3}) have the same order: 
coordinates, equations, and identities.

Substituting (\ref{g2}) and (\ref{g3}), we can readily verify that 
involutory condition (\ref{inv}) is satisfied.

The expressions (\ref{g2}) and (\ref{g3}) are also valid for 
the general case ($\tau\neq 0$), but for Eqs.\ (\ref{s1}) 
we must add in (\ref{g2}) the term ${}+ \delta ^{\rm i}_{n-2}$, 
since in this case the equation 
\[
{\bf e}_{{\rm nm}} - {\bf e}_{{\rm mn}}  =0 \ \ ({\rm m} ={\rm n}-1)
\]
in Eqs.\ (\ref{smb}) for $G^{{\rm n}-2}_2$ becomes an identity 
[in view of (\ref{gi2})]:
\be
{\bf e}_{{\rm nm}} - {\bf e}_{{\rm mn}} \sim S_{ {\rm nm} c,c} =0 \,;
\ee
here, $c\neq {\rm n,m}$ (we recall that $S_{abc}$ 
is completely antisymmetric tensor), and therefore this equation 
is contained in (\ref{gi2}).

We also give expressions for $\dim G_3^{\rm i}$
and $\dim G_4$, which are valid both for  (\ref{s1}) and for 
the general case:
\ba 
\dim G_3^{\rm i} & = & n^2 C^{n- {\rm i}+2}_3 - n^2 (n- {\rm i}+ 
n(1- \delta ^{\rm i}_n)\, ;\\
\dim G_3  & = & n^2 C^{n+4}_4 -n^3(n+1)/2 + n^2  \, .
\ea
Substituting these equations in condition (\ref{inv}), we can prove 
that the symbol $G_3$ is involutive.

4. In this section, we show that after the substitution
\be\label{sub}
h_a{}^\mu = H^{-p}H_a{}^\mu \,  \ \ (p=(1+1/\sigma -n)^{-1}; \ 
\ H= \det H^a{}_\mu \,; \ \ h= \det h^a{}_\mu)
\ee
systems (\ref{s1}) and (\ref{s2}) can be rewritten in a form that 
contains only the matrix $H_a{}^\mu $ (and its coordinate derivatives) 
but not $H^a{}_\mu $  ($H_a{}^\mu H^a{}_\nu = \delta^\mu_\nu $). 

Let us begin with Eq.\ (\ref{s2}), where $\sigma=0$ and hence
$H_a{}^\mu =h_a{}^\mu$. We rewrite (1) in the form
\be\label{l2}
\Lambda_{abc}= - h_{a\mu} 
(h_b{}^\mu{}_{,\nu}h_c{}^\nu - h_c{}^\mu{}_{,\nu}h_b{}^\nu)\,.
\ee
Using (\ref{l2}), we obtain in place of (\ref{s2}) the "trilinear" 
(by analogy with Hirota's bilinear form) equation:
\be\label{t2}
-h_a{}^\mu \Lambda_{abc,c} - \Lambda^{\mu}{}_{cd} \Lambda_{cdb} = 
(h_b{}^\mu{}_{,\nu}h_c{}^\nu - h_c{}^\mu{}_{,\nu}h_b{}^\nu)
{}_{,\lambda}h_c{}^{\lambda} + 
h_c{}^{\mu}{}_{,\lambda}
(h_c{}^\lambda{}_{,\nu}h_b{}^\nu - h_b{}^\lambda{}_{,\nu}h_c{}^\nu)
=0\,.
\ee

The next step is the 'unique system' (\ref{s1}), where $\sigma=1/3$ 
and $p=1/(4-n)$. Substituting (\ref{sub}) in (\ref{l2}) we obtain
\be\label{l3}
\Lambda_{abc}= - 2H^{-p}\left(
H_{a\mu }H_{[b}{}^{\nu } H_{c]}{}^{\mu }{}_{,\nu }
+p\eta _{a[b}H_{c]}{}^{\nu }A_{,\nu }\right),
\ee
\be 
\Phi _{c}= H^{-p}\left( H_{c}{}^{\nu}{}_{,\nu}
+{p\over \sigma}A_{,\nu }H_{c}{}^{\nu }\right)
\ \
(\mbox{here \ } A = \ln  H )\, . 
\ee
We also need the expression for $f_a{}^\mu$:
\be
f_{\epsilon\tau}= \Phi_{\epsilon,\tau} -\Phi_{\tau\epsilon}
=H_c{}^\lambda{}_{,\lambda\tau} H_{c\epsilon}+
H_c{}^\lambda{}_{,\lambda} H_{c\nu}
H_b{}^\nu{}_{,\epsilon} H_{b\tau} -  (\epsilon\tau)\,.
\ee

Rewriting Eq.\ (\ref{s1}) as follows
\be
H^{-p}\left[(-hL_a{}^{\mu\nu}){}_{,\nu}+\frac h 3 
(f_a{}^\mu + L_a{}^{\mu\nu}\Phi_\nu)\right]=0\,,
\ \ (h=H^{np+1})
\ee
we can ultimately reduce it (using the previous calculations)
to the trilinear form
\[ \left(\! H_a{}^\mu{}_{,\nu\lambda}H_b{}^\nu-
 H_b{\!}^\mu{}_{,\nu\lambda}H_a{}^\nu -
\frac23 H_a{}^\nu{}_{,\nu\lambda}H_b{}^\mu \!
\right) \! H_b{}^\lambda
+\frac13 H_b{}^\nu{}_{,\nu\lambda} \!
\left( H_a{}^\mu H_b{}^\lambda +
H_a{}^\mu H_b{}^\lambda \right) +
H_a{}^\mu{}_{,\nu}(H_b{}^\nu H_b{}^\lambda)_{,\lambda}
\]
\be \label{t1}
+ H_b{}^\mu{}_{,\nu}
(  H_b{}^\nu{}_{,\lambda}H_a{}^\lambda -
2 H_a{}^\nu{}_{,\lambda}H_b{}^\lambda -
H_b{}^\lambda{}_{,\lambda} H_a{}^\nu)
+\frac29 H_b{}^\lambda{}_{,\lambda} \!
\left(
H_b{}^\nu{}_{,\nu}H_a{}^\mu - H_a{}^\nu{}_{,\nu}H_b{}^\mu
\right)
=0 \, .
\ee

5. Examining trilinear equations (\ref{t2}) and (\ref{t1}), 
one may wonder whether the regularity of these equations 
is maintained at points of ${\cal R}_{2}$ 
at which the matrix $H_a{}^\mu$ is degenerate but finite.
If the symbol $G_2$ 
(or, in the case of 'unique equation' (\ref{t2}), the symbol $G_3$)
 remains involutive, i.e., the numbers $\dim G_2^{\rm i}$
($\dim G_3^{\rm i}$) do not change, then the answer 
will be in the affirmative.
Of course, we must not now multiply, for example, Eq.\ (\ref{t1}) by 
$H_a{}^\epsilon$, since some components of such system disappear, 
and there are changes in $\dim G_2$, $\dim G_3$, and so on, when
the matrix $H_a{}^\mu$ is degenerate, i.e., 
\[r = {\rm rank} H_a{}^\mu < n.
\]
In addition, for $r<n$ the order of the indices 
is important for the calculation of the values 
$\dim G_{\rm q}^{\rm i}$ (see [7]); 
it need not be changed if by choice of the coordinates 
we represent $H_a{}^\mu$ in the form
\be\label{hd}
H_a{}^\mu = {\rm diag}(0,\ldots,0,1,\ldots,1)= 
\delta_{\ul{a}}^\mu \ \ (=\delta_{\ul{b}}^\mu \delta^{\ul{b}}_a).
\ee
We use indices $\ul{a}, \ul{b} = n-r+1,\ldots,n$. 
Taking into account (\ref{hd}), one can write down an equation 
that determines the symbol
$G_2(r)$ for the trilinear system (\ref{t2}):
\be \label{vu1}
{\bf e}_{ab}= v_{ba,\ul{cc}}-v_{\ul{c}a,\ul{cb}} \,;
\ee
we change notation, $u_{ab} \to  v_{ba}$, to emphasize that now 
the 'working matrix' is $H_a{}^\mu$ (but not coframe matrix $h^a{}_\mu$).
Suppose $r=1$; then we obtain
\be
{\bf e}_{ab}= v_{ba,{\rm nn}}-v_{\rm{n}a,{\rm nn}}
\delta_b^{\rm n} = 0,\ \  
{\bf e}_{a{\rm n}} \equiv 0  \,;
\ee
here, $n$ is a fixed value of the index, and there is no
summation over $ n$.

It is clear that  $\dim G_2(r=1) > \dim G_2(r=n)$, and Eqs.\ (\ref{t2})
are irregular for $r=1$. It is easy to show that this conclusion
also holds for Eqs.\ (\ref{t1}). Therefore, it is not possible
to construct a formal solution in the form of a series beginning 
with $H_a{}^\mu(x_{(0)}^\nu)$ of $r=1$.

We now take $ r=2$ and in place of (\ref{vu1}) write
(we recall $m=n-1$; $m$ is also fixed value of index)
\be
{\bf e}_{ab}= v_{ba,{\rm nn}}+ v_{ba,{\rm mm}}
-v_{{\rm{}n}a,{\rm n}\ul{b}} - v_{{\rm{}m}a,{\rm m}\ul{b}}
 = 0  \,.
\ee
It can be shown that 
$ \dim G_2^{\rm i}(2) =
\dim G_2^{\rm i}(n)$: 
see Eq.\ (\ref{g2}) and note the identity of the last
equation with the equation for $G_2^{\rm i}(n)$.

It is somewhat more complicated to show for
'unique system' (\ref{t1}) that
$ \dim G_3^{\rm i}(2) =
\dim G_3^{\rm i}(n)$,
but it can be done.

The conclusion is that trilinear equations 
(\ref{t1}) and 
(\ref{t2})
\footnote{It turns out
that one-parameter class II${}_{221221}$
of EM-classification [1] comes via these two trilinear 
equations; all equations of this class (and that is all)
have trilinear presentation.
This 'trilinear class' has one 'common point'
with two-parameter class II${}_{22112}$,
but it has no common points with
two-parameter class I${}_{12}$
of lagrangian equations.}
are regular (involutive)
if $r\ge 2$.
At a point at which $r<n$, there will in general be
a singularity, since the scalars $\Lambda_{abc}$
become infinite.
It is true that, depending on $\sigma$, there will be restrictions 
on $n$. 
It is very interesting (and very important)
that for 'unique equation' and $n=5$ 
the minor of 'working matrix' 
(which is the frame density of some weight)
is equal to coframe matrix:
\be
\partial H^{-1}/\partial H_a{}^\mu
= H^{-1}H^a{}_\mu = h^a{}_\mu \, ;
\ee
perhaps, this equality makes impossible 
arising (contra)singularities in general
solutions (i.e., solutions of general position)
of 'unique equation' for space-time dimension
equal to five, because, as is shown in [8],
another sort of singularities 
(co-singularities with  rank$h^a_\mu \le n-1$)
is impossible (only) for the 'unique equation' (\ref{s1}).

Two difficult questions to AP (Pauli) about 'energy-momentum tensor' 
and about 'post-newtonian effects' are also considered in [8].
Some non-trivial topological issues of AP, 
topological classification of symmetrical
solutions and calculation of topological
quasi-charge groups for $n=5$ may be found in [9].


\end{document}